\documentstyle[graphicx,12pt]{article}
\begin{document}

\small
\hoffset=-1truecm
\voffset=-2truecm
\title{\bf The Casimir force on a piston in Randall-Sundrum models}
\author{Hongbo Cheng\footnote {E-mail address:
hbcheng@public4.sta.net.cn}\\
Department of Physics, East China University of Science and
Technology,\\ Shanghai 200237, China\\
The Shanghai Key Laboratory of Astrophysics,\\ Shanghai 200234,
China}

\date{}
\maketitle

\begin{abstract}
The Casimir effect of a piston for massless scalar fields which
satisfy Dirichlet boundary conditions in the context of
five-dimensional Randall-Sundrum models is studied. In these
scenarios we derive and calculate the expression for the Casimir
force on the piston. We also discuss the Casimir force in the
limit that one outer plate is moved to the remote place to show
that the nature of the reduced force between the parallel plates
left. In the Randall-Sundrum model involving two branes the two
plates attract each other when they locate very close, but the
reduced Casimir force turns to be repulsive as the plates
separation is not extremely tiny, which is against the
experimental phenomena. In the case of one brane model the shape
of the reduced Casimir force is similar to that of the standard
two-parallel-system in the four-dimensional flat spacetimes while
the sign of force remains negative.
\end{abstract}
\vspace{4cm} \hspace{1cm} PACS number(s): 03.70.+k, 11.10.Kk

\newpage

The various kinds of high-dimensional spacetimes have their own
compactification and properties of extra dimensions. The
approaches with additional dimensions were invoked for supplying a
breakthrough of the important problems such as the unification of
the interactions in nature, the hierarchy pullzle and the
cosmological constant etc.. In the Kaluza-Klein theory, one extra
dimension in our universe was introduced to be compactified to
unify gravity and classical electrodynamics [1, 2]. Some new
issues propose that the geometry of the extra spatial dimensions
is responsible for the hierarchy problem. First the large extra
dimensions (LED) were inserted [3-5]. In this approach the
additional dimensions are flat and of equal size and the radius of
a toroid is limited while the size of extra space can not be too
small, which can be used to solve the hierarchy problem. Second a
model with warped extra dimensions was also put forward by Randall
and Sundrum [6, 7]. The Randall-Sundrum (RS) model is a
five-dimensional theory compactiified on a $S^{1}/Z_{2}$ manifold
with bulk boundary cosmological constants leading to a stable
four-dimensional low energy effective theory suggesting that the
compact extra dimension with large curvature to explain the reason
why the large gap between the Planck and the electroweak scales
exists. In one of the RS models called RSI model, there are two
3-branes with equal opposite tensions and they are localized at
$y=0$ and $y=L$ respectively, with $Z_{2}$ symmetry
$y\longleftrightarrow-y$, $L+y\longleftrightarrow L-y$. The
Randall-Sundrum theory becomes RSII model as one brane is moved to
the infinity like $L\longrightarrow\infty$. The gauge fields
describing the standard model live on the negative tension brane
which is visible, while the positive tension brane with a
fundamental scale $M_{RS}$ is hidden.

The so-called Casimir effect is a remarkable macroscopic quantum
effect describing the attractive force between two conducting and
neutral parallel plates [8]. The physical manifestation of
zero-point energy receives great attention of the physical
community and has been extensively studied in a wide variety of
topics [9-15]. The precision of the measurement has been greatly
improved practically [16-19], leading the Casimir effect to be
remarkable observable and trustworthy consequence of the existence
of quantum fluctuations and to become a powerful tool for the
topics on the model of Universe with more that four dimensions.
The experimental results clearly show that the attractive Casimir
force between the parallel plates vanishes when the plates move
apart from each other to the very distant place. In particular it
should be pointed out that no repulsive Casimir force
appears.

A lot of efforts have been made to explore the high-dimensional
spacetimes by means of Casimir effect. Various kinds of models
with more than four dimensions are due to different
compactification of extra dimensions. Some topics related to the
Casimir effect have been studied extensive within the frame of
Kaluza-Klein theory. The research on the Casimir energy in the
context of Kaluza-Klein issue to explain the dark energy has been
performed and certainly is necessary, and more important results
have been obtained [20-22]. Having examined the Casimir effect for
the rectangular cavity in the presence of a compactified universal
extra dimension, we show analytically that the extra-dimension
influence on the original Casimir effect are very manifest [23].
The Casimir effect for the system of two parallel plates in the
spacetime with more than one additional compactified dimension was
discussed [24-27]. Further we show rigorously that there must
appear repulsive Casimir force between the parallel plates within
the experimental reach when the plates distance is sufficiently
large in this world possessing one compactified extra dimension
[25, 26]. We also generalize our investigation into the spacetimes
with more than one extra dimension within the frame of
Kaluza-Klein approach to show that the nature of Casimir force is
repulsive as the separation between the plates is large enough and
the more extra dimensions lead to the greater magnitude of
repulsive force [27]. In the spacetime with one additional
dimension compactified to a circle of radius $R$, the Casimir
force for an electromagnetic field between two perfectly
conducting parallel plates was also explored and it was shown that
there are two reflecting (discrete) modes and one penetrating
(continuum) mode propagating through the conductor. It is found
that they recover standard Casimir result in the limit as the
radius tends to vanish [28]. The UV cut-off dependence of the
Casimir energy for perfect conductors with an extra compact
dimension was also studied recently and the constraints from
Casimir effect experiments on the vacuum energy, the cosmological
constant and compact extra dimensions was obtained [29]. The
effect was probed extensively in the context of string theory
[30-32]. However the similar topics have also been investigated in
the braneworld. The cosmological constant can be interpreted as a
true Casimir effect for a scalar field filling the universe with
the topology [33]. It was shown that the vacuum energy vanishes
for one brane configuration and is not equal to zero for the case
of two branes [34]. It was also demonstrated that the nature of
the Casimir force can be repulsive in the Randall-Sundrum model
with two branes, which will change the radion stabilization
[35-37]. Recently some attention is paid to the Casimir effect for
two parallel plates [38-41]. The Casimir effect for the two
parallel plates in the five-dimensional Randall-Sundrum model was
discussed [38]. In the RSI model the constrains on the distance
between the two branes was obtained in virtue of the Casimir
effect for two-parallel-plate system and it is found that the
correction from warped factor to the Casimir force is too small to
be considered in the case of one brane [38]. The research has been
generalized to the case of high-dimensional Randall-Sundrum model
involving more compact dimensions and the similar conclusions were
drawn [39].

Recently some research on the Casimir effect for a new device
called piston in the backgrounds with more than four dimensions
have been performed. When we investigate the Casimir effect in a
confined region such as parallel plates, rectangular box and so
on, we can not neglect the contribution to the vacuum energy from
the area outside the surrounded region which has something to do
with the topology of the spacetime. The device of piston, a
slightly different model from the standard one, was first put
forward as a single rectangular box with dimensions $L\times b$
divided into two parts with dimensions $a\times b$ and
$(L-a)\times b$ respectively by a piston which is an idealized
plate that is free to move along a rectangular shaft [42]. Here
the author of Ref. [42] calculated the Casimir force on a
two-dimensional piston as a consequence of fluctuations of a
scalar field subject to Dirichlet boundary conditions on all
surfaces and showed that the forced on the piston is always
attractive as $L$ approaches to the infinity, regardless of the
ratio of the two sides. The Casimir force acting on a conducting
piston with arbitrary cross section remains attractive although
the existence of the walls weakens the force [43]. It was also
found that the sign of total Casimir force on a three-dimensional
piston for massless scalar fields obeying Dirichlet boundary
conditions is negative no matter how long the lengths of sides are
[44, 45]. In addition the Casimir force on a piston under various
boundary conditions may be repulsive [46]. The Casimir piston has
been studied in the environment with additional compactified
spatial dimensions. The Casimir force on the piston with perfect
magnetic conductor boundary conditions in general high-dimensional
spacetimes keeps attractive [47]. Within the frame of Kaluza-Klein
theory a Casimir piston for massless scalar fields satisfying
Dirichlet boundary conditions is probed. It is found that the
Casimir force affected by the extra dimensionality is attractive
between the piston and one plate in the limit that one outer plate
moved to the extremely distant place and more extra dimensions
gives rise to larger value of the attractive force [48, 49]. It is
further proved that the attractive Casimir force leads the center
plate to move towards the closer wall no matter how the
cross-section  of the piston and the geometry or topology of the
additional Kaluza-Klein dimensions are [50]. The thermal
corrections will not change the attractive nature of the Casimir
force [51].

It is necessary and significant to investigate the
force-on-the-piston problem in the Randall-Sundrum model.
According to the previous works a piston analysis is the correct
way to perform the parallel-plate calculation if the exterior
contributions can not be omitted. In fact we can obtain the same
form of Casimir force between the plates left as
$F_{C}=-\frac{\pi^{4}}{240}\frac{1}{a^{4}}$ in the limit that one
outer plate is moved to the remote place by means of the piston
analysis. Now we choose a piaton depicted in Fig. 1. As a piston,
one plate is inserted into a system consisting of two parallel
plates. The piston is parallel to the other plates and divides the
parallel-plate-system into two parts labelled by $A$ and $B$
respectively. In part $A$ the distance between the left plate and
the piston is $a$, and the distance between the piston and the
right plate in part $B$, the remains of the separation of two
plates, is certainly $L-a$, which means that $L$ denotes the whole
plates gap. We wonder the nature of the Casimir force between the
piston and the closer plate in the Randall-Sundrum scenario.
However little research has been devoted to this problem. The
intent of this letter is to consider the Casimir effect for the
system consisting of three parallel plates in the five-dimensional
RS model for simplicity and comparison to the conclusion of
conventional two-parallel-plate system. We regularize the force
expression arising with the help of the differential of the total
vacuum energy including the contribution outside the
three-parallel-plate model with respect to the distance between
two plates in the system to obtain the Casimir force on the piston
by means of zeta function technique as one outer plate is moved to
the remote place. It should be pointed out that we can make use of
the other regularization methods to obtain the Casimir force,
which means that our results are not regularization dependent.
Here we focus on the property of the Casimir force between one
plate and a piston in the RS models with one or two branes
respectively and compare our results from the piston models with
those of two parallel plates in the four-dimensional spacetime
supported by the experimental evidence. Our discussions and
conclusions are listed in the end.

We start to consider the massless scalar fields obeying Dirichlet
boundary conditions within the region involving a piston shown in
Fig. 1. The total vacuum energy for the three-parallel-plate
system can be written as the sum of three terms,

\begin{equation}
E=E^{A}(a)+E^{B}(L-a)+E^{out}
\end{equation}

\noindent where $E^{A}(a)$ and $E^{B}(L-a)$ mean the energy of
part $A$ and $B$ respectively, and the two terms depend on their
own size in parts. $E^{out}$ represents the vacuum energy outside
the system and is independent of characteristics inside the
device. Having regularized the total energy density, we can denote
the Casimir energy density as,

\begin{equation}
E_{C}=E_{R}^{A}(a)+E_{R}^{B}(L-a)+E_{R}^{out}
\end{equation}

\noindent where $E_{R}^{A}(a)$, $E_{R}^{B}(L-a)$ and $E_{R}^{out}$
are finite parts of terms $E^{A}(a)$, $E^{B}(L-a)$ and $E^{out}$
in Eq.(1) respectively. In particular, it is also pointed out that
the term $E_{R}^{out}$ is not a function of the position of the
piston, so the Casimir force on the piston is obtained by the
derivative of the Casimir energy with respect to the plates
distance like $-\frac{\partial E_{C}}{\partial a}$ and therefore
can be written as,

\begin{equation}
F'_{C}=-\frac{\partial}{\partial a}[E_{R}^{A}(a)+E_{R}^{B}(L-a)]
\end{equation}

\noindent which shows that the contribution of vacuum energy from
the exterior region does not affect the Casimir force on the
piston.

Here we set out to consider the massless scalar field living in
the bulk in the five-dimensional RSI model of spacetime. In this
issue the spacetime metric is chosen as,

\begin{equation}
ds^{2}=e^{-2k|y|}g_{\mu\nu}dx^{\mu}dx^{\nu}
\end{equation}

\noindent where the index runs from 0 to 3. Here the variable $k$
assumed to be of the order of the Planck scale governs the degree
of curvature of the $AdS_{5}$ with constant negative curvature.
Here the extra dimension compactified on an orbifold gives rise to
the generation of the absolute value of coordinate $y$ in metric
(4). Now we follow the procedure of Ref.[38, 39] to obtain the
dependence of a massless bulk scalar field denoted as
$\Phi(x^{\mu},y)$ on the variable $y$. In the five-dimensional
spacetime with the background metric shown in Eq.(1), the equation
of motion for the bulk scalar field reads
$\partial_{\mu}\partial^{\mu}\Phi+e^{2ky}\partial_{y}(e^{-4ky}\Phi)=0$
with usual four-dimensional flat metric with signature $-2$. The
field confining in the system consisting of three parallel plates
satisfies the Dirichlet boundary conditions
$\Phi(x^{\mu},y)|_{\partial\Omega}=0$, $\partial\Omega$ positions
of the plates expressed in space coordinates $x^{i}$, $i=1,2,3$,
then the wave vector in the directions restricted by the plates
may be $k_{n}=\frac{n\pi}{D}$, with $n$ a positive integer and $D$
the separation of the plates. We present the $y$-dependent part of
the bulk field $\Phi(x^{\mu},y)$ as $\chi^{(N)}(y)$ in virtue of
separation of variables. Having solved the equation of motion of
$\chi^{(N)}(y)$ we obtain the general expression of solutions for
the nonzero modes in terms of Bessel functions of the first and
second kind like
$\chi^{(N\neq0)}(y)=e^{2ky}[a_{1}J_{2}(\frac{m_{N}e^{ky}}{k})+a_{2}Y_{2}(\frac{m_{N}e^{ky}}{k})]$
with arbitrary constants $a_{1}$ and $a_{2}$. Here the effective
mass term for the bulk scalar field, $m_{N}$, can also be obtained
by means of integration out the fifth dimension $y$. In the case
of RSI model, the hidden and visible 3-branes are living at $y=0$
and $y=\pi R$ respectively, which leads to the modified Neumann
boundary conditions like $\frac{\partial\chi^{(N)}}{\partial
y}|_{y=0}=\frac{\partial\chi^{(N)}}{\partial y}|_{y=\pi R}=0$,
then a general reduced equation is shown as,

\begin{equation}
m_{N}\approx e^{-\pi kR}(N+\frac{1}{4})k\pi=\kappa(N+\frac{1}{4})
\end{equation}

\noindent where

\begin{equation}
\kappa=\pi ke^{-\pi kR}
\end{equation}

\noindent here we assume $N\gg1$ or equivalently $\pi kR\gg1$
throughout our work. Under these conditions, the zero-point
fluctuations of the fields can give rise to observable Casimir
forces among the plates and certainly the forces will display the
deviation from the properties of RSI model such as warp factor and
radion etc..

In Randall-Sundrum scenario we find that the frequency of the
vacuum fluctuation within a region confined by two parallel plates
with separation $D$ can be expressed as,

\begin{equation}
\omega_{nN}=\sqrt{k^{2}+(\frac{n\pi}{D})^{2}+m_{N}^{2}}
\end{equation}

\noindent where

\begin{equation}
k^{2}=k_{1}^{2}+k_{2}^{2}
\end{equation}

\noindent $k_{1}$ and $k_{2}$ are the wave vectors in directions
of the unbound space coordinates parallel to the plates surface.
Here $n$ and $N$ represent positive integers. Therefore the total
energy density of the bulk scalar fields in the interior of the
system involving two parallel plates in the RSI model reads,

\begin{eqnarray}
\varepsilon(D,\kappa)=\int
d^{2}k\sum_{n,N=1}^{\infty}\frac{1}{2}\omega_{nN}\hspace{7.5cm}\nonumber\\
=-\frac{\sqrt{\pi}}{4}\Gamma(-\frac{3}{2})[E_{2}(-\frac{3}{2};\frac{\pi^{2}}{D^{2}},
\kappa^{2};0,\frac{1}{4})-\kappa^{3}\zeta_{H}(-3,\frac{1}{4})-\frac{\pi^{3}}{D^{3}}
E_{1}(-\frac{3}{2};1;\frac{\kappa D}{4\pi})]
\end{eqnarray}

\noindent with the help of zeta functions of Epstein-Hurwitz type
such as
$E_{p}(s;a_{a},a_{2},\cdot\cdot\cdot,a_{p};c_{1},c_{2},\cdot\cdot\cdot,c_{p})
=\sum_{\{n_{j}\}=0}^{\infty}(\sum_{j=1}^{p}a_{j}(n_{j}+c_{j})^{2})^{-s}$,
$E_{1}(s;a;c)=\sum_{n=1}^{\infty}(an^{2}+c)^{-s}$ with $\{n_{j}\}$
a short notation of $n_{1},n_{2},\cdot\cdot\cdot,n_{p}$ and
$n_{j}$ a nonnegative integer and the Hurwitz zeta function like
$\zeta_{H}(s,q)=\sum_{n=0}^{\infty}(n+q)^{-s}$.

In the context of RSI model we investigate the
three-parallel-plate device depicted in Fig.1. Choosing the
variable $D=a$ in Eq.(9), we have the vacuum energy density for
part A of the system containing one plate and the piston with gap
$a$ as follow,

\begin{equation}
\varepsilon^{A}(a)=-\frac{\sqrt{\pi}}{4}\Gamma(-\frac{3}{2})[E_{2}
(-\frac{3}{2};\frac{\pi^{2}}{a^{2}},\kappa^{2};0,\frac{1}{4})
-\kappa^{3}\zeta_{H}(-3,\frac{1}{4})-\frac{\pi^{3}}{a^{3}}
E_{1}(-\frac{3}{2};1;\frac{\kappa a}{4\pi})]
\end{equation}

\noindent Similarly the vacuum energy density for the remains of
the system labelled B with plates distance $L-a$ by replacing the
variable $D$ with $L-a$ in Eq.(9) is,

\begin{eqnarray}
\varepsilon^{B}(L-a)\hspace{13cm}\nonumber\\
=-\frac{\sqrt{\pi}}{4}\Gamma(-\frac{3}{2})
[E_{2}(-\frac{3}{2};\frac{\pi^{2}}{(L-a)^{2}},\kappa^{2};0,\frac{1}{4})
-\kappa^{3}\zeta_{H}(-3,\frac{1}{4})-\frac{\pi^{3}}{(L-a)^{3}}
E_{1}(-\frac{3}{2};1;\frac{\kappa(L-a)}{4\pi})]
\end{eqnarray}

\noindent Following the procedures of Ref.[38], we regularize
Eq.(10) and Eq.(11) and then substitute the two regularized
expressions denoted as $\varepsilon_{R}^{A}(a)$ and
$\varepsilon_{R}^{B}(L-a)$ respectively into Eq.(3) to obtain the
Casimir force per unit area on the piston which divide the system
into two parts,

\begin{eqnarray}
f'_{IC}=-\frac{1}{2}\frac{\kappa^{2}}{a^{2}}\sum_{n_{1}=1}^{\infty}
\sum_{n_{2}=0}^{\infty}n_{1}^{-2}(n_{2}+\frac{1}{4})^{2}K_{2}
(2\kappa an_{1}(n_{2}+\frac{1}{4}))\hspace{4cm}\nonumber\\
-\frac{1}{2}\frac{\kappa^{3}}{a}\sum_{n_{1}=1}^{\infty}\sum_{n_{2}=0}^{\infty}
n_{1}^{-1}(n_{2}+\frac{1}{4})^{3}[K_{1}(2\kappa
an_{1}(n_{2}+\frac{1}{4}))+K_{3}(2\kappa
an_{1}(n_{2}+\frac{1}{4}))]\nonumber\\
+\frac{\pi}{32}\frac{\kappa^{2}}{a^{2}}\sum_{n=1}^{\infty}n^{-2}
K_{2}(\frac{\kappa
a}{2\sqrt{\pi}}n)+\frac{\sqrt{\pi}}{128}\frac{\kappa^{3}}{a}
\sum_{n=1}^{\infty}n^{-1}[K_{1}(\frac{\kappa a}{2\sqrt{\pi}}n)
+K_{3}(\frac{\kappa a}{2\sqrt{\pi}}n)]\hspace{0.3cm}\nonumber\\
+\frac{1}{2}\frac{\kappa^{2}}{(L-a)^{2}}\sum_{n_{1}=1}^{\infty}
\sum_{n_{2}=0}^{\infty}n_{1}^{-2}(n_{2}+\frac{1}{4})^{2}K_{2}
(2\kappa(L-a)n_{1}(n_{2}+\frac{1}{4}))\hspace{1.8cm}\nonumber\\
+\frac{1}{2}\frac{\kappa^{3}}{L-a}\sum_{n_{1}=1}^{\infty}
\sum_{n_{2}=0}^{\infty}n_{1}^{-1}(n_{2}+\frac{1}{4})^{3}\hspace{6.5cm}\nonumber\\
\times[K_{1}(2\kappa(L-a)n_{1}(n_{2}+\frac{1}{4}))+K_{3}(2\kappa(L-a)
n_{1}(n_{2}+\frac{1}{4}))]\hspace{1cm}\nonumber\\
-\frac{\pi}{32}\frac{\kappa^{2}}{(L-a)^{2}}\sum_{n=1}^{\infty}
n^{-2}K_{2}(\frac{\kappa(L-a)}{2\sqrt{\pi}}n)\hspace{5.8cm}\nonumber\\
-\frac{\sqrt{\pi}}{128}\frac{\kappa^{3}}{l-a}\sum_{n=1}^{\infty}
n^{-1}[K_{1}(\frac{\kappa(L-a)}{2\sqrt{\pi}}n)
+K_{3}(\frac{\kappa(L-a)}{2\sqrt{\pi}}n)]\hspace{3cm}
\end{eqnarray}

\noindent It is clear that Eq.(12) is the expression for the
Casimir force per unit area on the middle plate called piston
before the right plate of the system shown in Fig.1 has been moved
to the remote place. Further we take the limit
$L\longrightarrow\infty$ which means that the right plate in part
B is moved to a very distant place, then we obtain the following
expression for the Casimir force per unit area on the piston,

\begin{eqnarray}
f_{IC}=\lim_{L\longrightarrow\infty}f'_{IC}\hspace{10cm}\nonumber\\
=-\frac{1}{2}\frac{\kappa^{4}}{\mu_{I}^{2}}\sum_{n_{1}=1}^{\infty}
\sum_{n_{2}=0}^{\infty}n_{1}^{-2}(n_{2}+\frac{1}{4})^{2}K_{2}
(2\mu_{I}n_{1}(n_{2}+\frac{1}{4}))\hspace{4cm}\nonumber\\
-\frac{1}{2}\frac{\kappa^{4}}{\mu_{I}}\sum_{n_{1}=1}^{\infty}
\sum_{n_{2}=0}^{\infty}n_{1}^{-1}(n_{2}+\frac{1}{4})^{3}
[K_{1}(2\mu_{I}n_{1}(n_{2}+\frac{1}{4}))+K_{3}(2\mu_{I}n_{1}
(n_{2}+\frac{1}{4}))]\nonumber\\
+\frac{\pi}{32}\frac{\kappa^{4}}{\mu_{I}^{2}}\sum_{n=1}^{\infty}
n^{-2}K_{2}(\frac{\mu_{I}}{2\sqrt{\pi}}n)\hspace{7.5cm}\nonumber\\
+\frac{\sqrt{\pi}}{128}\frac{\kappa^{4}}{\mu_{I}}\sum_{n=1}^{\infty}
n^{-1}[K_{1}(\frac{\mu_{I}}{2\sqrt{\pi}}n)+K_{3}(\frac{\mu_{I}}{2\sqrt{\pi}}n)]
\hspace{4.8cm}
\end{eqnarray}

\noindent where

\begin{equation}
\mu_{I}=\kappa a=\pi kae^{-\pi kR}
\end{equation}

\noindent Here the parameter $\mu_{I}$ is dimensionless and the
connection between plates distance $a$ and the brane separation
$R$. In Eq.(13) $K_{\nu}(z)$ is the modified Bessel function of
the second kind and only the first several summands need to be
taken into account for numerical calculation to further discussion
because the terms with series converge very quickly. It is enough
to make the first several summands during the calculation. Having
performed the burden and surprisingly difficult calculation in
order to scrutinize the nature of the plate-piston Casimir force
in the RSI model, we show the dependence of the Casimir force on
the variable $\mu_{I}=\kappa a$ graphically in the Fig.2. It is
found that there must exist a special value denoted as
$\mu_{0}=0.156$. The sign of the Casimir force between one plate
and the piston is negative as $\mu_{I}<\mu_{0}$, meaning that the
two plates attract each other. When the distance between the plate
and piston is large enough like $\mu_{I}>\mu_{0}$, the nature of
the Casimir force keeps repulsive although the force vanishes as
the distance between the plates approaches to the infinity like
$\lim_{\mu_{I}\longrightarrow\infty}f_{IC}=0$. It is necessary to
make some estimations on the restriction on the gap between one
plate and the piston after the exterior plate is moved to the
extremely distant place in the case that we set $kR\sim12$ the
required value for solving the hierarchy puzzle and $k$ to be the
Planck scale. According to the definition of $\mu_{I}$ in Eq.(14)
and the special value $\mu_{0}=0.156$, we discover that the
Casimir force between plates is attractive only when the constrain
on plate separation is $a<2.2\times10^{-20}m$ or the force becomes
repulsive. We should not neglect that the repulsive Casimir force
between parallel plates is excluded in the practice within the
recent experiment reach. It should also be pointed out that this
experiment is always developed on the electromagnetic fields that
may obey more complicated boundary conditions than the case of
scalar field we consider here, but the shape of Casimir force
curves for different kinds of fields satisfying different boundary
conditions will be similar to the ones in Fig.2, which means that
the repulsive Casimir force must appear as the plate and piston
are sufficiently far away from each other. It should be pointed
out that the equation is valid asymptotically for $N\gg1$ although
the reduced equation (5) for the effective mass of the scalar bulk
field is expressed as an approximation. The error is about $0.03$
when $N=1$ and the error is $0.003$ and $0.001$ for $N=2$ and
$N=3$ respectively, etc., displaying that the error drops very
quickly with increasing $N$. The deviation from the approximation
is too weak to change our conclusion, so the repulsive Casimir
force denoted as positive magnitude of $f_{IC}$ will appear
inevitably when the plates distance is sufficiently large.

We proceed with the piston analysis in the case of RSII model with
one brane. In this issue the mass spectrum is continuous, so the
extra dimensional part of the mode summation in Eq.(9) will be
finished by means of an intefration with measure $\frac{dm}{k}$,

\begin{equation}
\varepsilon_{II}(D,k)=\frac{1}{2}\int d^{2}k\int\frac{dm}{k}
\sum_{n=1}^{\infty}\sqrt{k^{2}+(\frac{n\pi}{D})^{2}+m^{2}}
\end{equation}

\noindent The vacuum energy density for part A and B is equivalent
to that the variable $D$ of $\varepsilon(D)$ is chosen as $a$ and
$L-a$ respectively. Following the same procedures like zeta
function regularization and definition of Casimir force per unit
area on the piston in the background governed by RSII issue, we
obtain

\begin{equation}
f'_{IIC}=-\frac{\sqrt{\pi}}{2}\Gamma(\frac{5}{2})\zeta(5)\frac{1}{ka^{5}}
+\frac{\sqrt{\pi}}{2}\Gamma(\frac{5}{2})\zeta(5)\frac{1}{k(L-a)^{5}}
\end{equation}

\noindent where $\zeta(s)$ is the Riemann zeta function. In the
limit $L\longrightarrow\infty$, the Casimir force per unit area
between plates turns to be,

\begin{eqnarray}
f_{IIC}=\lim_{L\longrightarrow\infty}f'_{IIC}\hspace{1cm}\nonumber\\
=-\frac{\sqrt{\pi}}{4}\Gamma(\frac{5}{2})\zeta(5)\frac{1}{\mu_{II}^{5}}k^{4}
\end{eqnarray}

\noindent where

\begin{equation}
\mu_{II}=ka
\end{equation}

\noindent In fact this expression is equivalent to the relevant
results in Ref.[38]. The dependence of the Casimir pressure
between one and the piston living in the background described by
RSII model on the dimensionless variable $\mu_{II}$ defined in
Eq.(18) is depicted in Fig.3. The two plates keep on attracting
each other and the attractive Casimir force becomes weaker as the
plates are moved farther away from one another. The shape of the
expression of Casimir force per unit area in Eq.(17) is similar to
$f_{C}=-\frac{\pi^{4}}{240}\frac{1}{\mu_{II}^{4}}k^{4}$, the
Casimir force per unit plate area between two plates in the
four-dimensional flat spacetime. In Fig.4, we present the curve of
the absolute value of difference of the Casimir pressure in the
case of RSII scenario and the corresponding results in Minkowski
spacetime with four dimensions as $\Delta=|f_{IIC}-f_{C}|$
associated with the dimensionless parameter $\mu_{II}$ in unit of
$k^{4}$. We find that the magnitudes of the two kinds of force
will approach to the same when the distance between two plates is
very large, but as the plates gap is extremely tiny, their
difference will be very manifest. Within the region of the
required scales of curvature parameter $k$, certainly the product
$ka$ is very large, the plate-piston Casimir force in the context
of RSII model is equal to the standard results approximately.

By means of Casimir piston analysis the model of three parallel
plates in which the middle one is called piston is studied in the
presence of one warped additional dimension of the models put
forward by Randall and Sundrum in this work. It is discovered that
the five-dimensional Randall-Sundrum model with two branes can not
be realistic and the model in which the 3-brane at $y=\pi R$ is at
infinity may be acceptable. Here we also propose a new effective
method to explore the warped dimensions. According to the
definition of Casimir force its expression is obtained with the
help of zeta function regularization. When one outer plate is
moved away, we also get the reduced Casimir force per unit area
between one plate and the piston and the results about reduced
force are reliable because the exterior contribution has not been
omitted. In the limiting case we argue that there must appear the
repulsive Casimir force between the parallel plates in the
five-dimensional RSI model if the plate separation is not
extremely tiny and the magnitude of Casimir force tends to be aero
as the plate distance becomes large while the force nature remains
repulsive. The experimental evidence is against that the repulsive
Casimir force exists. Our analysis leads a reliable negative
verdict on the two-brane model. In the case of RSII model the
reduced Casimir force always keeps attractive. It is revealed that
the reduced force is similar to the force for the same piston
system in the four-dimensional flat spacetimes. The difference
between two kinds of forces can be negligible unless the plates
gap is infinitely small.

\vspace{3cm}

\noindent\textbf{Acknowledgement}

This work is supported by NSFC No.10875043 and is partly supported
by the Shanghai Research Foundation No.07dz22020.

\newpage
\begin{figure}
\setlength{\belowcaptionskip}{10pt} \centering
  \includegraphics[width=15cm]{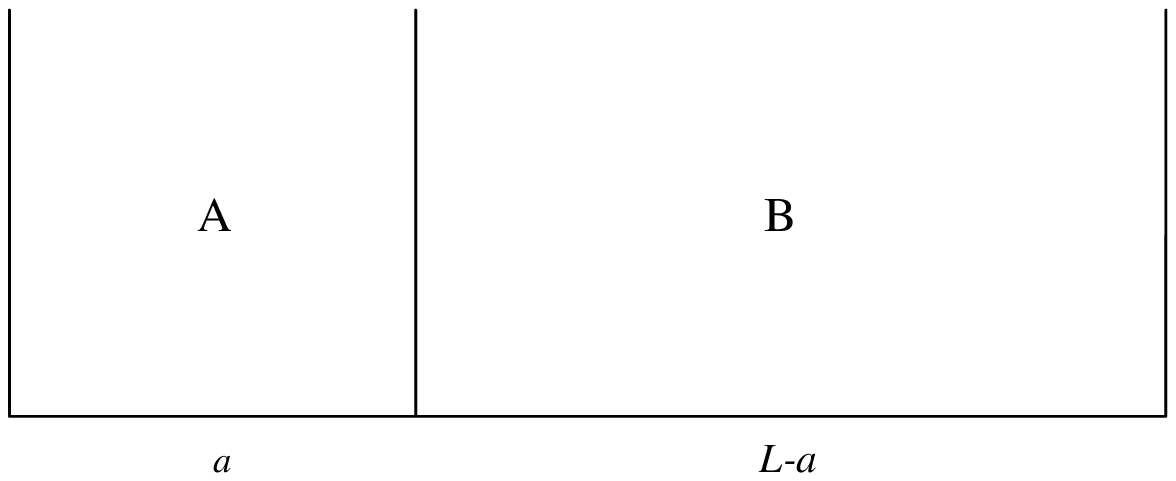}
  \caption{Casimir piston}
\end{figure}

\newpage
\begin{figure}
\setlength{\belowcaptionskip}{10pt} \centering
  \includegraphics[width=15cm]{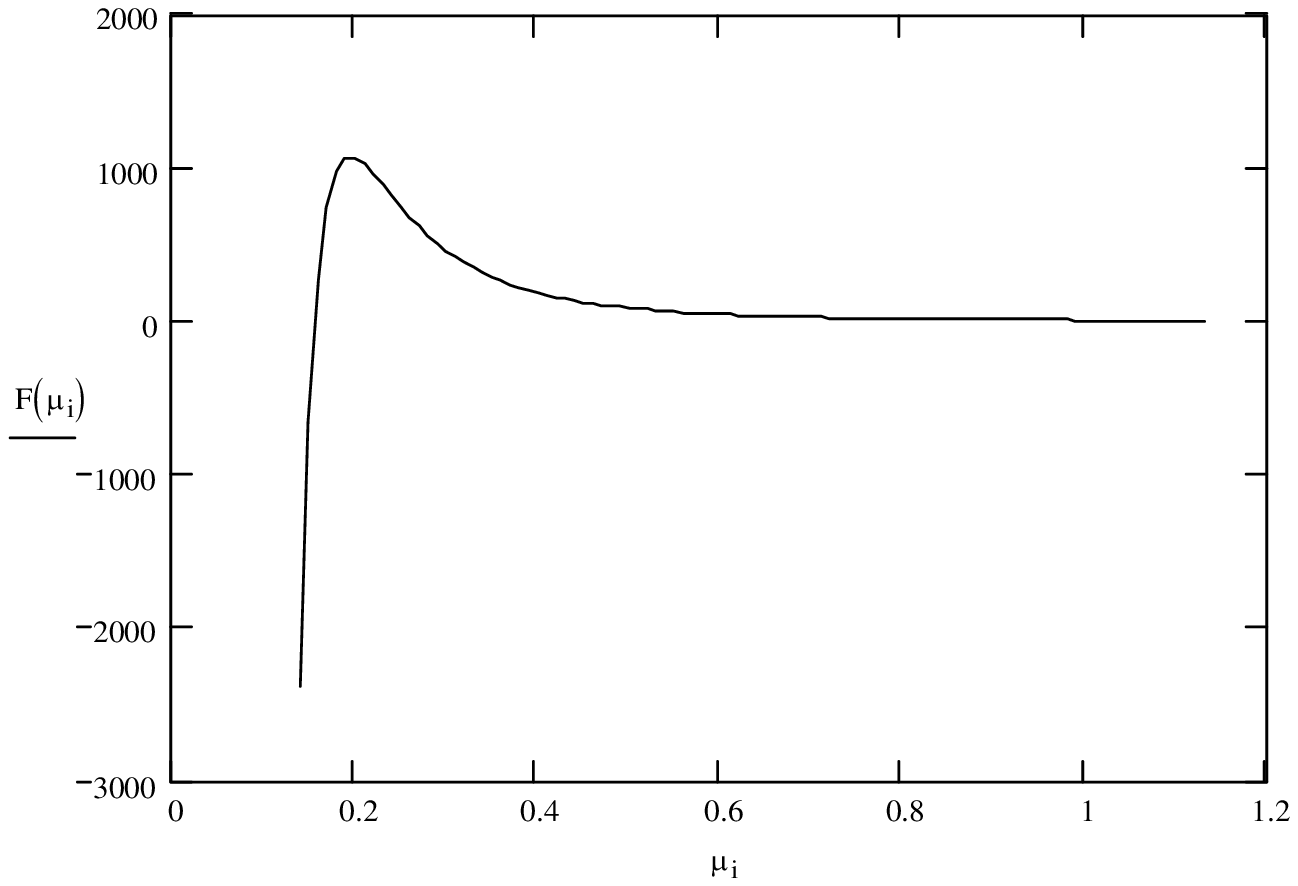}
  \caption{The reduced Casimir force per unit area in unit of $\kappa^{4}$
  between one plate and the piston versus the dimensionless variable $\mu_{RSI}=\kappa a$
  in the RSI model.}
\end{figure}

\newpage
\begin{figure}
\setlength{\belowcaptionskip}{10pt} \centering
  \includegraphics[width=15cm]{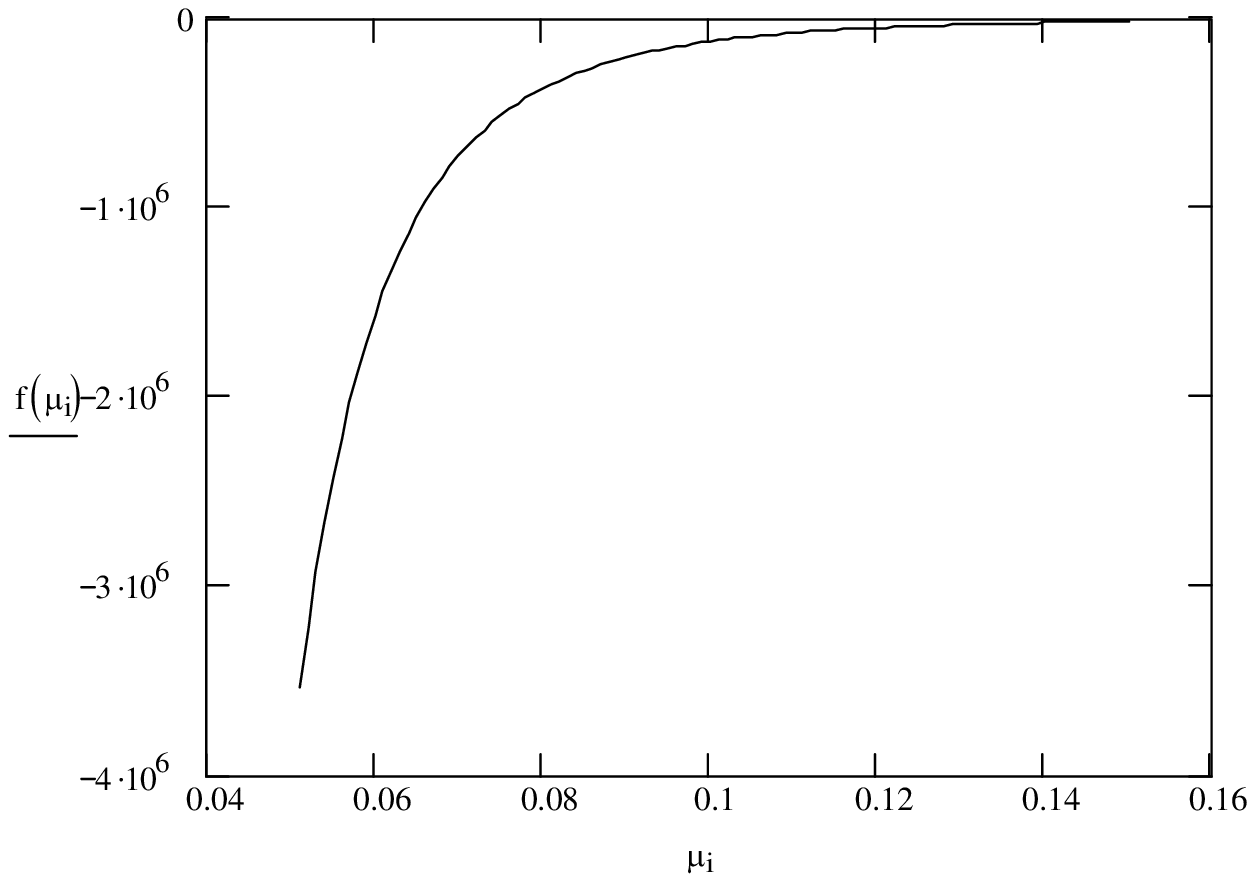}
  \caption{The reduced Casimir force per unit area in unit of $k^{4}$
  between one plate and the piston versus the dimensionless variable
  $\mu_{RSII}=ka$ in the RSII model.}
\end{figure}

\newpage
\begin{figure}
\setlength{\belowcaptionskip}{10pt} \centering
  \includegraphics[width=15cm]{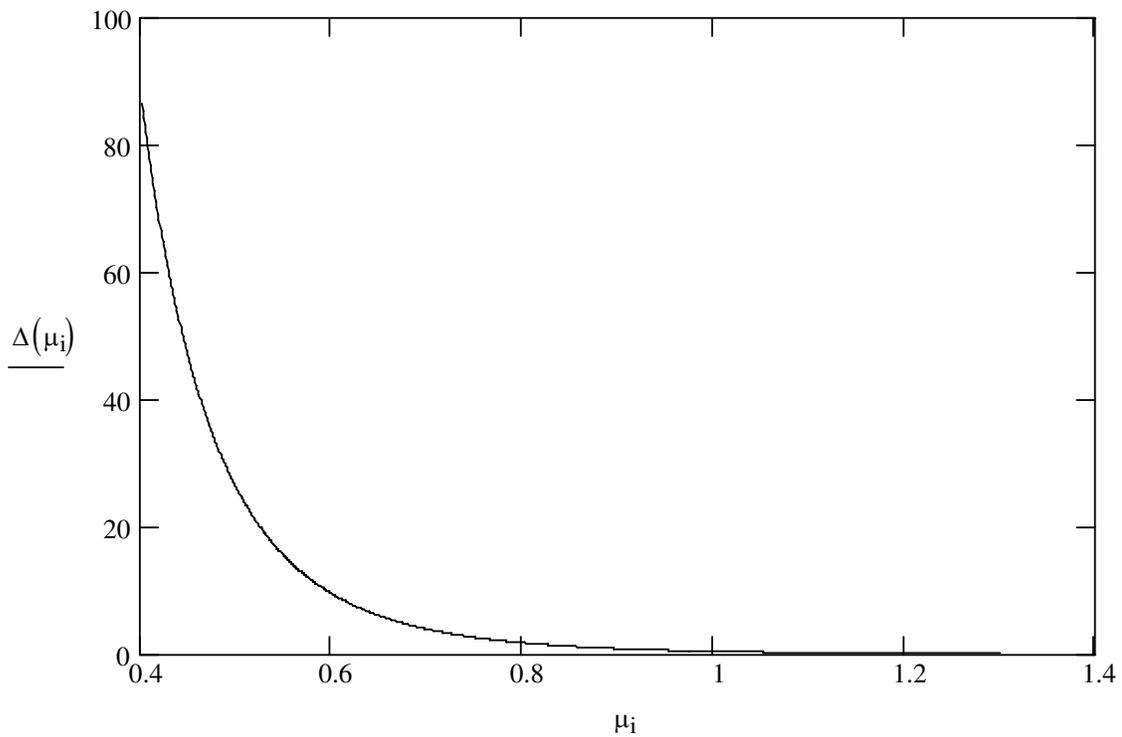}
  \caption{The absolute value of difference between the two reduced
  Casimir forces per unit area in unit of $k^{4}$ between one plate and the piston
  versus the dimensionless variable $\mu_{RSII}=ka$ in the RSII model and four-dimensional
  flat spacetime respectively.}
\end{figure}

\end{document}